\documentclass[10pt,twocolumn,letterpaper]{article}

\usepackage{wacv}
\usepackage{times}
\usepackage{epsfig}
\usepackage{graphicx}
\usepackage{amsmath}
\usepackage{amssymb}
\usepackage{amsthm}
\usepackage{caption}
\usepackage{subcaption}

\usepackage{url}
\usepackage[T1]{fontenc}

\usepackage{ctable} % for \specialrule command
\usepackage{bm}
\usepackage{amsmath,amsfonts,bm}

\def\eqref#1{equation~\ref{#1}}
\def\1{\bm{1}}

\def\vg{{\bm{g}}}

\def\vq{{\bm{q}}}

\def\vu{{\bm{u}}}

\def\vx{{\bm{x}}}
\def\vy{{\bm{y}}}

\DeclareMathAlphabet{\mathsfit}{\encodingdefault}{\sfdefault}{m}{sl}
\SetMathAlphabet{\mathsfit}{bold}{\encodingdefault}{\sfdefault}{bx}{n}

\newcommand{\E}{\mathbb{E}}

\newcommand{\R}{\mathbb{R}}

\newcommand{\Var}{\mathrm{Var}}

\DeclareMathOperator*{\argmax}{arg\,max}

\def\wacvPaperID{1035} % Enter the WACV Paper ID here

\wacvfinalcopy % *** Uncomment this line for the final submission

\ifwacvfinal
\fi

\ifwacvfinal
\usepackage[breaklinks=true,bookmarks=false]{hyperref}
\else
\usepackage[pagebackref=true,breaklinks=true,colorlinks,bookmarks=false]{hyperref}
\fi

\begin{document}

%%%%%%%%% TITLE
\title{On the Effectiveness of Small Input Noise for Defending Against Query-based Black-Box Attacks}
\author{Junyoung Byun, Hyojun Go, Changick Kim\\
Korea Advanced Institute of Science and Technology (KAIST)\\
{\tt\small \{bjyoung, gohyojun15, changick\}@kaist.ac.kr}}

\maketitle

\ifwacvfinal
\thispagestyle{empty}
\fi

%%%%%%%%% ABSTRACT
\begin{abstract}
While deep neural networks show unprecedented performance in various tasks, the vulnerability to adversarial examples hinders their deployment in safety-critical systems. Many studies have shown that attacks are also possible even in a black-box setting where an adversary cannot access the target model's internal information. Most black-box attacks are based on queries, each of which obtains the target model's output for an input, and many recent studies focus on reducing the number of required queries. In this paper, we pay attention to an implicit assumption of query-based black-box adversarial attacks that the target model's output exactly corresponds to the query input. If some randomness is introduced into the model, it can break the assumption, and thus, query-based attacks may have tremendous difficulty in both gradient estimation and local search, which are the core of their attack process. From this motivation, we observe even a small additive input noise can neutralize most query-based attacks and name this simple yet effective approach Small Noise Defense (SND). We analyze how SND can defend against query-based black-box attacks and demonstrate its effectiveness against eight state-of-the-art attacks with CIFAR-10 and ImageNet datasets. Even with strong defense ability, SND almost maintains the original classification accuracy and computational speed. SND is readily applicable to pre-trained models by adding only one line of code at the inference.
\end{abstract}

%%%%%%%%% BODY TEXT

\begin{figure*}
 \begin{subfigure}[b]{0.65\textwidth}
         \centering
         \includegraphics[height=4.3cm,trim={0cm 0.3cm 0cm 0.3cm},clip]{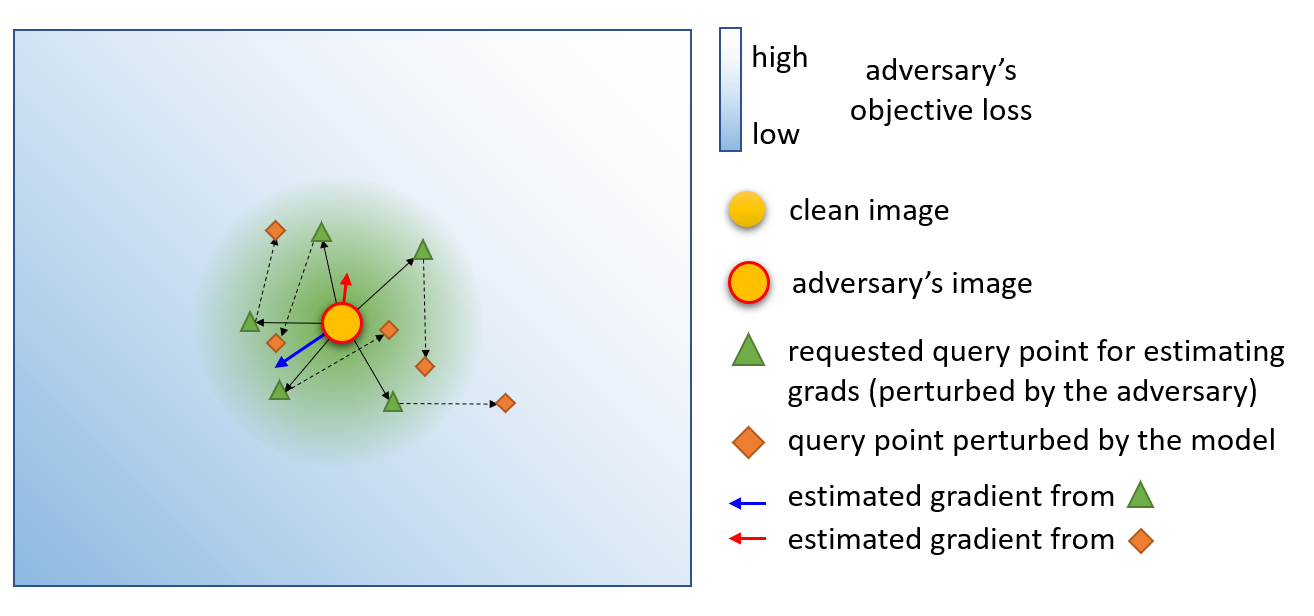}
         \caption{}
     \end{subfigure}
     \begin{subfigure}[b]{0.35\textwidth}
         \centering
         \includegraphics[height=4.3cm,trim={0cm 0.0cm 0cm 0cm},clip]{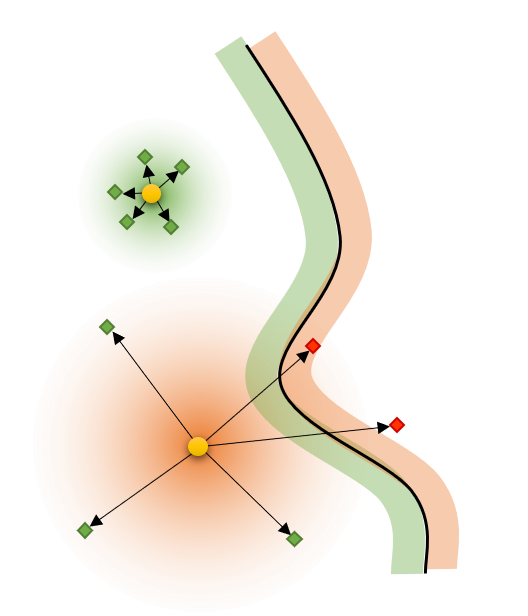}
         \caption{}
     \end{subfigure}
        \caption{Illustrations of our intuitions. (a) Small noise can effectively disturb gradient estimation of query-based attacks which use finite difference. (b) Compared to large noise, small noise hardly affects predictions on clean images.}
        \label{fig:fig1}

\end{figure*}

\maketitle

\section{Introduction}
Although deep neural networks perform well in various areas, it is now well-known that small and malicious input perturbation can cause them to malfunction \cite{biggio2013evasion,szegedy2013intriguing}. This vulnerability of AI models to adversarial examples hinders their deployment, especially in safety-critical areas. In the white-box setting, where the target model's parameters can be accessed, strong adversarial attacks such as Projected Gradient Descent (PGD) \cite{madry2018towards} can generate adversarial examples using the internal information. Recent studies have shown that adversarial examples can be generated even in a practical black-box setting where the model's interior is hidden to adversaries.

These black-box attacks can be largely divided into \textit{transfer-based attacks} and \textit{query-based attacks}. Transfer-based attacks train a substitute model that mimics the target model's behavior and take advantage of \textit{transferability} that adversarial examples generated from a network can deceive other networks \cite{papernot2017practical}. However, due to differences in training methods and model architectures, the transferability of adversarial examples can be significantly weakened, and thus, transfer-based attacks usually result in lower success rates \cite{chen2017zoo}. For this reason, most black-box attacks are based on \textit{queries}, each of which obtains the target model's output for an input.

Query-based attacks create adversarial examples through an iterative process based on either local search with repetitive small input modifications or optimization with estimated gradients of an adversary's loss with respect to an input. Here, requesting many queries in their process takes a lot of time and financial loss. Moreover, many similar query images can be suspicious to system administrators. For this reason, researchers have focused on reducing the number of queries required to make a successful adversarial example \cite{bhambri2019survey}.

Compared to the increasing number of studies on adversarial defenses in white-box settings, the number of defenses against query-based black-box attacks is still very small \cite{bhambri2019survey}. However, in a practical situation, black-box attacks are more realistic as attackers cannot know the target model's interiors. Since existing defenses developed for white-box attacks improve their robustness at the high cost of clean accuracy (accuracy on clean images) \cite{tsipras2018robustness}, it is necessary to develop a new defense strategy that targets query-based black-box attacks with minimal accuracy loss.

To defend against query-based black-box attacks, we pay attention to an implicit but important assumption of these attacks that \emph{the target model's output exactly corresponds to the query input}. If some randomness is introduced into the model, it can break the assumption, and thus, query-based attacks may have tremendous difficulty in both gradient estimation and local search, which are the core of their attack process. This intuition is illustrated in Fig. \ref{fig:fig1}(a).

Among previous studies, Dong \textit{et al.} \cite{dong2020benchmarking} empirically find that randomization-based defenses are more effective in defending against query-based black-box attacks than other types of defenses. However, existing randomization-based defenses introduce significant uncertainty into predictions, and thus, they also degrade clean accuracy. 

In this paper, however, we highlight that simply adding small Gaussian noise into an input image is enough to defeat various query-based attacks by breaking the above core assumption while almost maintaining clean accuracy. One may think that additive Gaussian noise cannot defend against most adversarial attacks unless we introduce large randomness. This idea is valid for white-box attacks \cite{gu2014towards}, but we will show that small noise is surprisingly effective against query-based black-box attacks.

Our second intuition is that sufficiently small Gaussian noise hardly affect predictions on clean images, as shown in Fig. \ref{fig:fig1}(b). Dodge \textit{et al.} \cite{dong2020benchmarking} empirically find that classification accuracy decreases in proportion to the variance of Gaussian noise, but the accuracy drop is negligible for a sufficiently small variance. 

We think an adversarial defense techniques should have the following goals:
(1) preventing malfunction of a model against various attacks, (2) minimizing the computational overhead, (3) maintaining the accuracy on clean images, and (4) easily applicable to existing models. The proposed defense against query-based attacks meets all of the above objectives, and we name this simple yet effective defense technique \textit{Small Noise Defense} (SND).

Our contributions can be listed as follows:
\setlength{\leftmargini}{0.5cm}
\begin{itemize}
\setlength\itemsep{-0.1em}
  \item We highlight the effectiveness of adding a small additive noise to input for defending against query-based black-box attacks. The proposed SND method can be readily applied to pre-trained models by adding only one line of code in the PyTorch framework \cite{NEURIPS2019_9015} at the inference stage (\verb|x = x + sigma * torch.randn_like(x)|) and almost maintains the performance of the model.
  \item We analyze how SND can efficiently interfere with gradient estimation and local search, which are the core of query-based attacks.
  \item We devise an adaptive attack method against SND and explain its limitations and the difficulty of evading SND.
  \item We have empirically shown that the proposed method can effectively defend against eight state-of-the-art query-based black-box attacks with the CIFAR-10 and ImageNet datasets. Specifically, four decision-based and four score-based attacks are used to show strong defense ability against various attacks, including local search-based and optimization-based methods.
\end{itemize}

\section{Background}
\subsection{Adversarial Setting}
Since we deal with adversarial attacks on the image classification task throughout the paper, we briefly explain adversarial attacks on the image classification task in this section.

Suppose that a neural network $f(\vx)$ classifies an image $\vx$ among total $N$ classes and returns a class-wise probability vector $\vy = [y_1, ..., y_N]$ for $\vx$. For notational convenience, we also denote the probability of $i^{th}$ class (i.e., $y_i$) as $f(\vx)_i$ and the top-1 class index as $h(\vx) = \argmax_{i \in C} y_i$, where $C=\{1, ..., N\}$.

In a black-box threat model, an adversary has a clean image $\vx_0$ whose class index is $c_0$ and wants to generate an adversarial example $\hat{\vx} = \vx_0 + \bm{\delta}$ to fool a target model $f$. In the following, we denote the adversarial example at $t^{th}$ step in an iterative attack algorithm as $\hat{\vx}_t = \bm{x}_0+\bm{\delta}_t$. The adversary should generate an adversarial example within a perturbation norm budget $\epsilon$ and query budget $Q$. If we let $q$ be the number of queries used to make $\bm{\delta}_t$, then we can write the adversary’s objective as follows:
\begin{equation}
    \min_{\bm{\delta}_t}\ell(\vx_0+\bm{\delta}_t)\textrm{, } 
    \text{subject to } ||\bm{\delta}_t||_p\leq \epsilon \text{ and } q \leq Q,
\end{equation}
where $\ell(\vx)=f(\vx)_{c_0}-\max_{c \neq c_0}f(\vx)_c$ for untargeted attacks and $\ell(\vx)=\max_{c \neq \hat{c}}f(\vx)_c-f(\vx)_{\hat{c}}$ for targeted attacks with target class index $\hat{c}$.
Since decision-based attacks cannot obtain $\ell(\vx)$, they have a different objective for untargeted attacks as follows:
\begin{equation}
    \min_{\bm{\delta}_t}||\bm{\delta}_t||_p\textrm{, } 
    \text{subject to } h(\bm{x}_0)\neq h(\bm{x}_0+\bm{\delta}_t) \text{ and } q \leq Q.
\end{equation}
Unless otherwise noted, in this paper, we use $p=2$ and focus on untargeted attacks because it is more challenging for defenders. Besides, we assume that each pixel value is normalized into $[0, 1]$.

\subsection{Taxonomy of query-based black-box attacks}
Query-based attacks can be largely divided into score-based and decision-based attacks according to the available type of output of the target model (class-wise probabilities for score-based attacks and the top-1 class index for decision-based attacks).
On the other hand, query-based attacks can be categorized into optimization-based attacks and local search-based attacks. Optimization-based methods optimize an adversary's objective loss with estimated gradients of the loss with respect to $\hat{\vx}_t$. In contrast, local search-based attacks repeatedly update an image according to how the model's output changes after adding a small perturbation. 

In the following, we briefly introduce various query-based attacks used in this paper.

\textbf{Bandit optimization with priors (Bandit-TD).} Ilyas \textit{et al.} \cite{ilyas2018prior} observe that the image gradients in successive steps of an iterative attack have strong correlations. In addition, they find that the gradients of surrounding pixels also have correlations. Bandit-TD exploits this information as priors for efficient gradient estimation.

\textbf{Simple Black-box Attack (SimBA \& SimBA-DCT).}
For each iteration, SimBA \cite{guo2019simple} samples a vector $\vq$ from a pre-defined set Q and modify the current image $\hat{\vx}_t\textrm{ with }\hat{\vx}_t-\vq$ and $ \hat{\vx}_t+\vq$ and updates the image in the direction of decreasing $\vy_{c_0}$. Inspired by the observation that low-frequency components make a major contribution to misclassification \cite{guo2018low}, SimBA-DCT exploits DCT basis in low-frequency components for query-efficiency.

\textbf{Boundary Attack (BA).} BA \cite{brendel2018decision} updates $\hat{\vx}_t$ on the decision-boundary so that the perturbation norm gradually decreases via random walks while misclassification is maintained.

\textbf{Sign-OPT.} Cheng \textit{et al.} \cite{cheng2018query} treat a decision-based attack as a continuous optimization problem of the nearest distance to the decision boundary. They use the randomized gradient-free method \cite{nesterov2017random} for estimating the gradient of the distance. Cheng \textit{et al.} \cite{cheng2019sign} propose Sign-OPT, which uses the expectation of the sign of gradient with random directions to estimate the gradients efficiently without exhaustive binary searches.

\textbf{Hop Skip Jump Attack (HSJA).}
Chen \textit{et al.} \cite{chen2020hopskipjumpattack} improve BA with gradient estimation. For each iteration of HSJA, it finds an image on the boundary with a binary search algorithm, and estimates the gradients, and calculates the step-size towards the decision boundary.

\textbf{GeoDA.}
Rahmati \textit{et al.} \cite{rahmati2020geoda} propose a geometry-based attack that exploits a geometric prior that the decision boundary of the neural network has a small curvature on average near data samples. By linearizing the decision boundary in the vicinity of samples, it can efficiently estimate the normal vector of the boundary, which helps to reduce the number of required queries for generating adversarial examples.

\subsection{Adversarial Defenses}
As Dong \textit{et al.} \cite{dong2020benchmarking} observe that randomization is important for effective defense against query-based attacks, we focus on randomization-based defenses among various defense methods. In what follows, we briefly explain three randomization-based defenses along with PGD-adversarial training.

\textbf{Random Self-Ensemble (RSE).} RSE \cite{liu2018towards} adds Gaussian noise with $\sigma_\textrm{inner}=0.1$ to the input of each convolutional layer, except for the first convolutional layer where $\sigma_\textrm{init}=0.2$ is used. To stabilize the performance, they use an ensemble of multiple predictions for each image.

\textbf{Parametric Noise Injection (PNI).}
He \textit{et al.} \cite{he2019parametric} propose a method to increase the robustness of neural networks by adding trainable Gaussian noise to the activation or weight of each layer. They introduce learnable scale factors of noise and allow them to be learned with adversarial training.

\textbf{Random Resizing and Padding (R\&P).}
Xie \textit{et al.} \cite{xie2018mitigating} propose a random input transform-based method. In front of network inference, it applies random resizing and random padding to its input sequentially, making adversaries obtain noisy gradients. It can be easily applied to a pre-trained model, but it increases total computational time due to the enlarged input image.

\textbf{PGD-Adversarial Training (PGD-AT).} Madry \textit{et al.} \cite{madry2018towards} propose PGD-adversarial training which trains a model with adversarial examples generated by Projected Gradient Descent (PGD) \cite{madry2018towards}. Unlike other defenses that are ineffective against adaptive attacks, it is well known that PGD-AT provides strong defense against a variety of white-box attacks.
\section{Analysis}
\subsection{Our Approach}
To defend against query-based black-box attacks, we add Gaussian noise with a sufficiently small $\sigma$ to the input as follows.
\begin{equation}
    f_{\bm{\eta}}(\vx)=f(\vx+\bm{\eta})\textrm{, where }\bm{\eta} \sim \mathcal{N}(\bm{0},\sigma^2\bm{I})\textrm{ and }\sigma\ll1.
\end{equation}
For an adversary, since the exact value of $\bm{\eta}$ is unknown, there are multiple possible output values for any $\vx$, so $f_{\bm{\eta}}$ is a random process. In what follows, we will explain how this transform introduces tremendous difficulty in both gradient estimation and local search in query-based black-box attacks.

\subsection{Defense Against Optimization-based Attacks}\label{sec_opt}
In this subsection, we will explain how small Gaussian input noise can disturb the gradient estimation in optimization-based attacks. We first look at defense against score-based attacks and then deal with decision-based attacks.

The core of optimization-based attacks is an accurate estimation of $\nabla\ell(\vx)$, which needs to be approximated with finite difference because of the black-box setting. For instance, the gradient can be estimated as $\Tilde{\vg}$ by Random Gradient-Free method \cite{nesterov2017random} as follows.
\begin{equation}
\begin{gathered}
    \Tilde{\vg}=\frac{1}{B}\sum_{i=0}^B\bm{g}_i, \\\textrm{where } \bm{g}_i=\frac{\ell(\hat{\vx}_t+\beta \vu) - \ell(\hat{\vx}_t)}{\beta}\vu\textrm{ and }\vu \sim \mathcal{N}(\bm{0},\sigma^2\bm{I}).
\end{gathered}
\label{eqn:gradient}
\end{equation}
Conceptually, by introducing small Gaussian noise into input, $\Tilde{\vg}$ can greatly differ from the true gradient $\nabla\ell$ as shown in Fig. \ref{fig:fig1}.

To illustrate it more formally, let us represent $\bm{\eta}$ by replacing it with $\bm{\eta}(\vx)$ to clarify $\bm{\eta}$ depends on both time and $\vx$. Suppose $f_{\bm{\eta}(\vx)}^*$ is a sample function of the random process $f_{\bm{\eta}(\vx)}(\vx)$ at some time. Then, this function is noisy with regard to $\vx$ because of $\bm{\eta}(\vx)$. We also assume that $\ell^*$ is derived from $f_{\bm{\eta}(\vx)}^*$, then unless $\Var[\ell^*(\vx+\vu)]$ is extremely small, $\ell^*$ is discontinuous and non-differentiable, and thus, $\nabla\ell^*$ does not exist. Therefore, the estimated gradient using finite differences does not converge to the target gradient $\nabla\ell$. For example, simplifying the problem, if $f$ is a one-dimensional function $\R \rightarrow \R$ and sampled $\eta(1.000)=0.08$ and $\eta(1.001)=-0.03$ then, the sampled function values $f(x+\eta(x))$ at $x=1.000$ and $x=1.001$ become $f(1.080)$ and $f(0.9701)$, respectively. Therefore, if the variance of $f$ is large, the sample function becomes more noisy.

\setlength{\tabcolsep}{4pt}
\begin{table*}[t]

\centering
\begin{tabular}{lclc}
\specialrule{.1em}{.05em}{.05em} 
\multicolumn{2}{c}{ResNet-20 on CIFAR-10} & \multicolumn{2}{c}{ResNet-50 on ImageNet} \\
Defense & \multicolumn{1}{l}{Clean Accuracy (\%)} & Defense & \multicolumn{1}{l}{Clean Accuracy (\%)} \\ \hline
Baseline & 91.34 & Baseline & 76.13 \\
SND ($\sigma =0.001$) & 91.33 $\pm$ 0.02 & SND ($\sigma =0.001$) & 76.10 $\pm$ 0.02 \\
SND ($\sigma =0.01$) & 90.57 $\pm$ 0.09 & SND ($\sigma =0.01$) & 75.47 $\pm$ 0.03 \\
SND ($\sigma =0.02$) & 87.56 $\pm$ 0.18 & SND ($\sigma =0.02$) & 73.91 $\pm$ 0.02 \\
RSE & 83.40 $\pm$ 0.15 & PGD-AT & 57.9 \\
PNI & 85.15 $\pm$ 0.18 & R\&P & 74.26 $\pm$ 0.07 \\
\specialrule{.1em}{.05em}{.05em} 
\end{tabular}
\caption{Comparison of clean accuracy. For randomization-based methods, we denote the mean and standard deviation of clean accuracy in 5 repetitive experiments with different random seeds.}
\label{table:clean_acc}
\end{table*}

In decision-based attacks, $\hat{\vx}_t$ is likely to be in the vicinity of the decision boundary. Therefore, even small noise can move $\hat{\vx}_t$ across the boundary so that the output is changed. The estimated gradient through erroneous predictions hinders the generation of adversarial examples. We illustrate the working principle of SND against decision-based attacks in supplementary material. Besides, the binary search algorithm, which is widely used to calculate the distance to the decision boundary \cite{chen2020hopskipjumpattack,cheng2019sign}, can make a larger error due to $\bm{\eta}$. Therefore, algorithms such as HSJA, which assume that $\hat{\vx}$ is near the decision boundary, are likely to work incorrectly.

\subsection{Defense Against Local Search-based Attacks}
Local search-based attacks try to update the image in the direction that decreases the adversarial objective loss. However, since the output is unreliable due to noise, this becomes similar to random motion. Suppose an adversary recognizes that the attack objective loss decreases for $\hat{\bm{x}}_{t}+\bm{q}$, where $\bm{q}$ is a perturbation, and updates $\hat{\bm{x}}_{t+1}$ as $\hat{\bm{x}}_{t}+\bm{q}$. However, since the actually evaluated input of $f$ is $\hat{\bm{x}}_{t}+\bm{q}+\bm{\eta}$,  where $\bm{\eta} \sim \mathcal{N}(\bm{0},\sigma^2\bm{I})$, the attack objective loss might increase at the originally intended input $\hat{\bm{x}}_{t}+\bm{q}$. This prediction error makes the attack algorithm stuck in the iterative process and prevents generating adversarial examples.

\subsection{Adaptive Attacks on Small Noise Defense} \label{sec_diff}
Solid research for adversarial defense requires evaluating the defense ability against adaptive attacks that exactly know the working principle of the defense \cite{carlini2019evaluating}.
Athalye \textit{et al.} \cite{athalye2018obfuscated} show that randomization-based defense such as R\&P can be circumvented through the Expectation Over Transform (EOT) technique \cite{athalye2018synthesizing}. The EOT technique approximates the gradient at each gradient descent step by averaging gradients w.r.t. several samples transformed by randomization-based defenses. In the black-box setting, the gradients w.r.t. an input cannot be obtained at once, and it should be estimated using the finite-difference by giving several queries. Therefore, EOT-based adaptive attacks in black-box settings average outputs for each input to get a reliable prediction for accurate gradient estimation.

The above way of averaging the noisy output (i.e., expectation) is one of two primary techniques for handling noise in derivative-free optimization \cite{wang2018noisy}. The other primary technique is threshold selection which stores a reliable candidate solution set (in our framework, candidate adversarial examples) that makes minimal losses. It accepts a new solution for the candidate set when the evaluated loss is less than the recorded smallest loss by the threshold. This principle of threshold selection can be adapted to the gradient estimation step in query-based attacks. Since decision-based attacks try to reduce the size of adversarial perturbation, selecting a candidate set with a sufficiently large threshold is identical to increasing $\sigma$ of random perturbations in the gradient estimation like Eq. \ref{eqn:gradient} to ignore the disturbance of small input noise. However, increasing the perturbation size can amplify the gradient estimation error by itself.

Therefore, following the suggestion of \cite{athalye2018obfuscated,wang2018noisy}, we design expectation-based adaptive attacks against SND. In the following, we shed light on the difficulty of evading the proposed defense with the adaptive attacks in detail. 

In our framework, the input of the function, $\vx+\bm{\eta}$, is a Gaussian random process. But the result of the nonlinear function $f$, $f(\vx+\bm{\eta})$, is no longer a Gaussian random process. This makes it very difficult for query-based attacks to bypass SND. For expectation-based adaptive attacks, adversaries can approximate $f(\vx)$ as $\E_{\bm{\eta}}[f_{\bm{\eta}}(\vx)]$ by taking the average over multiple queries using the fact that $\E(\bm{\eta})=\bm{0}$. However, this attempt requires many queries for each iteration and greatly diminishes query efficiency. We note that adversaries should also consider the query efficiency for their adaptive attacks.

In addition, even if a large amount of queries are used, $\E_{\bm{\eta}}[f_{\bm{\eta}}(\vx)]$ may be different from $f(\vx)$ because of the nonlinearity of the deep neural networks.
With simple examples, we will explain how the expectation value differs from the actual value when Gaussian noise is added to the input of nonlinear functions.

\setlength{\tabcolsep}{4pt} % Default value: 6pt

\begin{table*}[t]

\centering
\begin{tabular}{lcccccccccccc}
\specialrule{.1em}{.05em}{.05em} 
Attack method & \multicolumn{3}{c}{BA} & \multicolumn{3}{c}{Sign-OPT} & \multicolumn{3}{c}{HSJA} & \multicolumn{3}{c}{GeoDA} \\
\# of queries & \multicolumn{1}{c}{2K} & \multicolumn{1}{c}{5K} & \multicolumn{1}{c}{10K} & \multicolumn{1}{c}{2K} & \multicolumn{1}{c}{5K} & \multicolumn{1}{c}{10K} & \multicolumn{1}{c}{2K} & 5K & 10K & \multicolumn{1}{c}{2K} & \multicolumn{1}{c}{5K} & 10K \\ \hline
Baseline & 36.2\ & 69.5\ & 84.6\ & 59.1\ & 88.9\ & 91.2\ & 86.4\ & 89.2\ & 89.2\ & 64.7\ & 71.3\ & 76.5\ \\
SND ($\sigma =0.01$) & 14.0\ & 17.8\ & 20.0\ & 21.7\ & 22.3\ & 22.8\ & 16.5\ & 19.9\ & 22.7\ & 12.0\ & 12.1\ & 12.4\ \\
SND ($\sigma =0.001$) & 33.0\ & 53.0\ & 61.3\ & 20.6\ & 22.2\ & 23.4\ & 48.1\ & 67.6\ & 81.9\ & 11.7\ & 11.8\ & 12.2\ \\
RSE & 18.2\ & 19.0\ & 19.8\ & 18.7\ & 18.7\ & 18.7\ & 19.9\ & 22.2\ & 23.4\ & 19.7\ & 19.9\ & 20.5\ \\
PNI & 15.1\ & 15.4\ & 15.8\ & 18.1\ & 18.1\ & 18.1\ & 17.4\ & 19.2\ & 20.6\ & 19.9\ & 20.2\ & 20.4\ \\ 
\specialrule{.1em}{.05em}{.05em} 
\end{tabular}
\caption{Evaluation of attack success rates (\%) on the CIFAR-10 dataset.}
\label{table:cifar10}
\end{table*}
\begin{table*}[t]
\centering

\begin{tabular}{lccccccccc}
\specialrule{.1em}{.05em}{.05em} 
Attack type & \multicolumn{9}{c}{Decision-based Attack} \\ \hline
Attack method & \multicolumn{3}{c}{Sign-OPT} & \multicolumn{3}{c}{HSJA} & \multicolumn{3}{c}{GeoDA} \\ 
\# of queries & 5K & 10K & 20K & 5K & 10K & 20K & 5K & 10K & 20K \\ \hline
Baseline & 36.4\% & 62.4\% & 88.0\% & 64.0\% & 88.4\% & 99.6\% & 50.0\% & 62.8\% & 72.0\% \\
 & {(}9.71{)} & {(}4.72{)} & {(}2.38{)} & {(}4.43{)} & {(}2.34{)} & {(}1.28{)} & {(}6.38{)} & {(}5.04{)} & {(}4.12{)} \\
SND ($\sigma{=}0.01$) & \textbf{6.8\%} &\textbf{6.8\%} & \textbf{7.2\%} & \textbf{6.4\%} & \textbf{7.6\%} & \textbf{8.4\%} & \textbf{7.2\%}& \textbf{7.6\%} & \textbf{7.6\%} \\
 & {(}41.37{)} & {(}70.27{)} & {(}87.94{)} & {(}31.03{)} & {(}26.81{)} & {(}22.93{)} & {(}33.48{)} & {(}33.14{)} & {(}32.61{)} \\
SND ($\sigma{=}0.001$) &\textbf{6.8\%} & 7.6\% & 8.0\% & 13.6\% & 20.4\% & 32.4\% & 8.4\% & 8.8\% & 9.2\% \\
 & {(}35.15{)} & {(}54.33{)} & {(}88.29{)} & {(}11.60{)} & {(}8.86{)} & {(}6.75{)} & {(}27.44{)} & {(}25.88{)} & {(}24.41{)} \\
PGD-AT & 28.8\% & 30.0\% & 32.4\% & 30.4\% & 33.2\% & 36.0\% & 32.4\% & 34.0\% & 35.6\% \\
 & {(}30.67{)} & {(}24.66{)} & {(}19.72{)} & {(}20.99{)} & {(}17.27{)} & {(}13.46{)} & {(}14.54{)} & {(}13.63{)} & {(}12.90{)} \\
R\&P & 13.2\% & 13.2\% & 13.2\% & 13.6\% & 15.2\% & 16.0\% & 14.4\% & 14.4\% & 15.2\% \\
 & {(}51.19{)} & {(}82.14{)} & {(}85.78{)} & {(}33.01{)} & {(}31.00{)} & {(}29.42{)} & {(}31.72{)} & {(}31.21{)} & {(}30.45{)} \\ 
\specialrule{.1em}{.05em}{.05em} 
Attack type & \multicolumn{9}{c}{Score-based Attack} \\\hline
Attack method & \multicolumn{3}{c}{SimBA} & \multicolumn{3}{c}{SimBA-DCT} & \multicolumn{3}{c}{Bandit-TD} \\
\# of queries & 5K & 10K & 20K & 5K & 10K & 20K & 5K & 10K & 20K \\ \hline
Baseline & 74.0\% & 74.4\% & 74.4\% & 94.8\% & 95.2\% & 95.2\% & 94.0\% & 97.2\% & 98.4\% \\
 & {(}3.89{)} & {(}3.99{)} & {(}4.02{)} & {(}3.12{)} & {(}3.14{)} & {(}3.14{)} & {(}4.70{)} & {(}4.70{)} & {(}4.70{)} \\
SND ($\sigma{=}0.01$) & \textbf{8.4\%} & \textbf{9.2\%} & \textbf{10.0\%} & \textbf{8.4\%} & \textbf{8.8\%} & \textbf{10.4\%} & 15.2\% & 15.2\% & 16.4\% \\
 & {(}0.52{)} & {(}0.55{)} & {(}0.57{)} & {(}0.56{)} & {(}0.58{)} & {(}0.60{)} & {(}4.74{)} & {(}4.74{)} & {(}4.74{)} \\
SND ($\sigma{=}0.001$) & 27.2\% & 35.6\% & 50.4\% & 46.4\% & 60.4\% & 68.4\% & \textbf{7.2\%} & \textbf{7.6\%} & \textbf{8.4\%} \\
 & {(}1.84{)} & {(}2.14{)} & {(}2.43{)} & {(}2.22{)} & {(}2.44{)} & {(}2.58{)} & {(}4.83{)} & {(}4.83{)} & {(}4.83{)} \\
PGD-AT & 27.6\% & 27.6\% & 27.6\% & 36.0\% & 36.0\% & 36.0\% & 38.8\% & 45.2\% & 52.8\% \\
 & {(}5.46{)} & {(}7.55{)} & {(}10.17{)} & {(}5.36{)} & {(}6.21{)} & {(}6.62{)} & {(}3.54{)} & {(}3.54{)} & {(}3.54{)} \\
R\&P & 26.4\% & 27.6\% & 28.0\% & 27.2\% & 28.4\% & 29.2\% & 32.0\% & 33.2\% & 33.6\% \\
 & {(}0.48{)} & {(}0.51{)} & {(}0.54{)} & {(}0.52{)} & {(}0.55{)} & {(}0.58{)} & {(}4.50{)} & {(}4.50{)} & {(}4.50{)} \\ \specialrule{.1em}{.05em}{.05em} 
\end{tabular}
\caption{Evaluation of attack success rates against defenses on the ImageNet dataset. We denote the average $\ell_2$ norm of perturbations in the parenthesis.}
\label{table:imagenet}
\end{table*}
\newtheorem{example}{Example}
\begin{example}[A simple nonlinear function]
\label{example1}
Let $F(\vx)=\vx^T\vx$, where $F:\R^d\rightarrow\R$, and $F_{\bm{\eta}}(\vx)=F(\vx+\bm{\eta})\textrm{ where }\bm{\eta} \sim \mathcal{N}(\bm{0},\sigma^2\bm{I})$. Suppose we estimate $F(\bm{0})$ with $\E[F_{\bm{\eta}}(\bm{0})]$. Then, $\E[F_{\bm{\eta}}(\bm{0})]=\E[(\bm{0}+\bm{\eta})^T(\bm{0}+\bm{\eta})]=\E[\bm{\eta}^T\bm{\eta}]=d\sigma^2$. Therefore, $\E[F_{\bm{\eta}}(\bm{0})]=d\sigma^2 \neq 0 = F(\bm{0})$ and if $d$ is very large (e.g., for an image of size $224{\times}224{\times}3$, $d{=}150,528$), then the estimation error would be high.

\end{example}
\begin{example}[A simple ReLU network case]
\label{example2}
Let $\text{ReLU}(x)$ $=\text{max}(0,x)$ and $F(x) = \text{ReLU}(wx + b)$,  where $F: \R \xrightarrow{} \R$. Let $F_\eta(x) = F(x+ \eta)$, where $\eta$ $\sim \mathcal{N}(0,\sigma^2)$ and suppose we estimate $F(x)$ with $\E [F_\eta(x)]$. Then $\E [F_\eta(x)]$ is as follows:
\begin{equation}
         (wx+b)(1 - \Phi\left(-\frac{wx+b}{|w|\sigma}\right)) + |w|\sigma\phi\left(-\frac{wx+b}{|w|\sigma}\right).
\end{equation}
\end{example}
\begin{proof}
Let $w(x+\eta) + b$ be $Y$, then $Y$ can be represented with $\mu_y = wx + b$ and $\sigma_y^2 = w^2\sigma^2$ as:
\begin{equation}
    Y \sim \mathcal{N}(\mu_y, \sigma_y^2).
\end{equation}
Then, $F_\eta(x)$ is $\text{max}(0,Y)$ and $\E[F_\eta(x)]=\E[\text{max}(0,Y)]$ can be obtained by the law of total expectation.
\begin{equation}
\begin{aligned}
    \E[F_\eta(x)] & = \E[\text{max}(0,Y)] \\ 
    &= \E[Y|Y>0]\mathrm{Pr}(Y>0) + 0\mathrm{Pr}(Y\leq0). \\ 
\end{aligned}
\end{equation}
Using the truncated normal distribution, we recall the fact as follows:
\begin{equation}
    \E[Y|Y > a] = \mu_y + \sigma_y \frac{\phi((a-\mu_y)/\sigma_y)}{1-\Phi((a-\mu_y)/\sigma_y)},
\end{equation}
where $\phi(x) = \frac{1}{\sqrt{2\pi}}\text{exp}(-\frac{1}{2}x^2)$ and $\Phi(\cdot)$ is the cumulative distribution function of the standard normal distribution. Since $\mathrm{Pr}(Y>0))= 1-\Phi(\frac{\mu_y}{\sigma_y})$, $\E[F_\eta(x)]$ is represented as:
\begin{equation}
\begin{aligned}
        & \mu_y  (1- \Phi\left(-\frac{\mu_y}{\sigma_y}\right)) + \sigma_y \phi \left(-\frac{\mu_y}{\sigma_y}\right)\\
&= (wx+b)(1 - \Phi\left(-\frac{wx+b}{|w|\sigma}\right)) + |w|\sigma\phi\left(-\frac{wx+b}{|w|\sigma}\right).
\end{aligned}
\end{equation}
Therefore, if the noise is added to the input in the simple ReLU case, there can be a difference between the actual $F(x)$ value and the estimated value by $\E[F_\eta(x)]$. 
\end{proof}
From the proof on the simple network, we can expect that the average of the output may have an error with the actual output even in a deep neural network.

\section{Experiments and Discussion}
\subsection{Experimental Settings}
In this section, we evaluated the defense ability of SND against eight  query-based black-box attacks: BA, Sign-OPT, HSJA, GeoDA, SimBA, SimBA-DCT, Bandit-TD, and Subspace Attack, along with other defense methods: PNI, RSE, R\&P, and PGD-AT. We used the CIFAR-10 \cite{krizhevsky2009learning} and ImageNet \cite{imagenet_cvpr09} datasets for our experiments and following previous studies \cite{chen2020hopskipjumpattack, guo2019simple, he2019parametric}, we used ResNet-20 for CIFAR-10 and ResNet-50 \cite{he2016identity} for ImageNet as target networks. Following \cite{brendel2018decision}, we randomly sampled 1,000 and 250 correctly classified images from the CIFAR-10 test set and the ImageNet validation set for evaluation. We describe detailed experimental settings in supplementary material.

For evaluation metrics, we first define a \textit{successfully attacked image} as an image from which an attack can find an adversarial image within the perturbation budget $\epsilon$ and query budget $Q$. With this definition, we use \textit{attack success rate}, which is the percentage of the number of successfully attacked images over the total number of evaluated images. We note that since we evaluate defense performance, \textbf{a lower attack success rate is better}. We measured the $\ell_2$ norm of perturbations and set $\epsilon$ to $1.0$ for the CIFAR-10 dataset, and to $5.0$ for the ImageNet dataset. Note that we denote the $q^{th}$ query image as $\hat{\vx}^q$. $\hat{\vx}^q$ and $\hat{\vx}_t$ can be different.

\subsection{Evaluation of Clean Accuracy}
We first evaluated the clean accuracy of models with defenses on the original test split (10K images) of the CIFAR-10 and validation split (50K images) of the ImageNet dataset.
As shown in Table \ref{table:clean_acc}, SND hardly reduces the clean accuracy compared to other methods. The accuracy drop caused by SND is not significant at $\sigma\leq0.01$, which implies that sufficiently small $\sigma$ hardly affects clean accuracy.

\subsection{Evaluation on the CIFAR-10 Dataset}
We performed four decision-based attacks against models with defenses, and Table \ref{table:cifar10} shows the evaluated attack success rates. SND shows competitive defense ability despite having more than 5\% higher clean accuracy compared to other defenses. Moreover, due to significant performance drop, RSE and PNI cannot be applied to the models for large-scale image classification with the ImageNet dataset.

\setlength{\tabcolsep}{3pt} % Default value: 6pt

\subsection{Evaluation on the ImageNet Dataset}
We performed six query-based attacks against models with defenses, and Table \ref{table:imagenet} shows the evaluated attack success rates. When the query budget $Q$ is 20K, the average of the attack success rates over the attacks against the baseline is 87.9\%, whereas SND with $\sigma=0.01$ significantly reduces it to 10.0\%. SND with $\sigma=0.001$ also significantly reduces the average attack success rate to 29.5\%, which is comparable to the second-best method, R\&P (22.5\%).

We also calculated the average $\ell_2$ norm of perturbations of query images $||\vx_0-\hat{\vx}^q||_2$ at the predefined query budget $Q$ to show whether the perturbation norm diverges or not. If an attack stops in the middle without requesting $Q$ queries, we used the last query image instead. In decision-based attacks, it can be seen that randomization-based defenses, SND and R\&P, significantly increase the perturbation norm as $q$ increases. In SimBA and SimBA-DCT, the perturbation norm is minimal in SND and R\&P, which implies that the attacks have significant difficulty in finding a perturbation which decreases $y_{c_0}$.

\subsection{Empirical Evidence for Assumptions of SND}
To provide supporting evidence for assumptions of SND in score-based attacks, we calculated $\hat{\sigma}=\sqrt{\Var[\ell(\vx+\bm{\eta})]},$ where $\bm{\eta} \sim \mathcal{N}(\bm{0},\sigma^2\bm{I})$. With $\sigma=0.01$ and 1,000 test images of the CIFAR-10, we evaluated $\ell(\vx+\bm{\eta})$ for 100 iterations for each clean image and calculated the $\hat{\sigma}$ averaged over all images. In our experiment, $\hat{\sigma}=0.04$ which is small but sufficient to make $\ell(\vx+\bm{\eta})$ to be non-differentiable about $\vx$.

In decision-based attacks, our assumption in Section \ref{sec_opt} is that if $\hat{\vx}_t$ is near the decision boundary, small noise can easily move the image across the boundary. We evaluated $P_{mis}:=P(h(\vx)\neq h(\vx+\bm{\eta}))$ through experiments. We counted the above mismatch case for all queries during the attack process. 
With $\sigma=0.001\textrm{, }0.01$ and the CIFAR-10 test images, the average of $P_{mis}$ over all the attacks is calculated as 0.22 and 0.25, respectively. In contrast, on clean images, $P_{mis}$ is obtained as 0.002 and 0.021, respectively. Therefore, the results shown in Table \ref{table:prob} support our argument.

\subsection{Evaluation of Adaptive Attacks Against SND}
As described in Section \ref{sec_diff}, we devised an adaptive attack against SND that takes the expectation of predictions for repetitive $T$ queries. In this experiment, we performed HSJA against SND with $\sigma=0.01$ on the CIFAR-10 dataset. Since HSJA is a decision-based attack, we regard the most predicted class in $T$ queries as the expected class. We measured the attack success rate and $P_{mis}$ according to the query budget, and the adaptive attack clearly shows a higher attack success rate than the baseline ($T=1$), as shown in Table \ref{table:diff}. On the same query budget, however, the adaptive attack shows a lower attack success rate (e.g., 22.7\% ($T{=}1$) > 18.2\% ($T{=}5$) at $Q$=10K, and
29.3\% ($T{=}5$) > 27.3\% ($T{=}10$) at $Q$=50K). Therefore, the expectation-based adaptive attack has limitations due to the restricted query budget. Moreover, even if $T$ increases, $P_{mis}$ does not decrease and this reinforces our argument in Section \ref{sec_diff}.  
We also applied the adaptive attack to BA, SO, and GeoDA on CIFAR-10 and two score-based attacks (SimBA-DCT and Bandit-TD) on ImageNet for comprehensive comparisons. The experimental results are shown in  supplementary material.

\begin{table}[t]
    \centering
        
        \begin{tabular}{lcc}
        \specialrule{.1em}{.05em}{.05em} 
        \multicolumn{1}{r}{Defense} & SND & SND \\
        Attack  & ($\sigma=0.001$) & ($\sigma=0.01$) \\ \hline
        BA & 0.134 & 0.227 \\
        Sign-OPT & 0.216 & 0.215 \\
        HSJA & 0.255 & 0.189 \\
        GeoDA & 0.314 & 0.391 \\
        None & 0.002 & 0.021 \\ \specialrule{.1em}{.05em}{.05em} 
        \end{tabular}
        \caption{Evaluation of $P(h(\vx)\neq h(\vx+\bm{\eta}))$.}
        \label{table:prob}
\end{table}
\begin{table}[t]
    \centering
        
        \begin{tabular}{lcccc}
        \specialrule{.1em}{.05em}{.05em} 
        \# of queries & 2K$\times T$ & 5K$\times T$ & 10K$\times T$ & $P_{mis}$\\\hline
        HSJA   ($T{=}1$) & 16.5\% & 19.9\% & 22.7\% & 0.189 \\
        HSJA   ($T{=}5$) & 18.2\% & 23.8\% & 29.3\% & 0.321 \\
        HSJA   ($T{=}10$) & 21.0\% & 27.3\% & 34.9\% & 0.376 \\
        HSJA   ($T{=}20$) & 25.2\% & 34.0\% & 46.6\% & 0.410 \\\hline
        \# of queries & 50K & 100K & 200K &  \\
        HSJA  & 29.0\% & 30.1\% & 31.0\% & 0.120\\ 
        \specialrule{.1em}{.05em}{.05em}
        \end{tabular}
        \caption{Attack success rates of the adaptive version of HSJA against SND ($\sigma=0.01$) with different $T$.}
        \label{table:diff}
\end{table}

\subsection{Varying $\sigma$ for Each Inference} % OK
So far, we have used a fixed $\sigma$ for SND.
Changing $\sigma$ for each query may reduce clean accuracy while maintaining the defense ability. From this motivation, we multiplied $\bm{\eta}$ with $k$ which is randomly sampled between 0 and 1 from the beta distributions with three different settings: (1) Uniformly random (the same as $\alpha{=}\beta{=}1$) (2) Sampling from a beta distribution with $\alpha{=}\beta{=}2$ whose probability density function (PDF) is $\cap$-shaped. (3) Sampling from a beta distribution with $\alpha{=}\beta{=}0.5$ whose PDF is $\cup$-shaped. We calculated clean accuracy and average $\ell_2$ norm of noise for each method. Among the three ways, at $\sigma=0.01$, $\alpha{=}\beta{=}2$ is better than the others in terms of the loss of clean accuracy and the defensive ability against Sign-OPT. Detailed results can be found in supplementary material.% 

\subsection{Defense Against Hybrid Black-box Attacks} % OK
Since SND protects models by interfering with gradient estimation and the local search of query-based attacks, we do not expect SND is effective against transfer-based attacks as these attacks exploit the transferability of adversarial examples. However, our method is complementary with other defenses, which are mainly effective against transfer-based black-box attacks such as \cite{jalwana2020orthogonal} and \cite{tramer2018ensemble}. Combined with other defense techniques, SND can work well against general black-box attacks, provided that the model's parameters are kept secret to adversaries.

To support this argument, we experimented with Subspace attack \cite{guo2019subspace}, a hybrid attack that exploits transferability in query-based attacks. Specifically, the Subspace attack exploits transferability-based priors, which are gradients from local substitute models trained on a small proxy dataset. We used pre-trained ResNet-18 and ResNet-34 \cite{he2016identity} as reference models for gradient priors. We performed the attack based on the $\ell_\infty$ norm because the authors provide parameter settings only for the $\ell_\infty$ norm.

Detailed experimental results are described in supplementary material, but the results show that SND alone cannot effectively defend against the hybrid attack with gradient priors. However, when SND is combined with PGD-AT, it effectively protects the model and decreases the attack success rate from 100\% to 42.4\% at $Q$=20K and $\sigma{=}0.01$. To focus on the defensive ability against gradient estimation, we recalculated the attack success rate without initially misclassified images. Then, the newly obtained attack success rate decreases from 100\% to 16.4\% at $Q$=20K. This result implies that SND can be combined with other defenses against transfer-based attacks to achieve strong defense ability against all types of black-box attacks.

\section{Related Work}
The idea of injecting random noise at the inference stage for improving adversarial robustness is not new \cite{pinot2019theoretical,pinot2021robustness,nemcovsky2019smoothed}.
However, we note that small input noise, which is insufficient to prevent white-box attacks, is surprisingly effective in defending models against query-based black-box attacks.

\subsection{History-based Detection Methods Against Query-based Black-box Attacks.}
To the best of our knowledge, studies that mainly target defending against query-based black-box attacks have not yet been published. However, history-based detection techniques for query-based attacks have been proposed recently \cite{chen2019stateful,li2020blacklight}. Considering that adversary requires many queries of similar images for finding an adversarial example, they store information about past query images to detect the unusual behavior of query-based attacks.

\subsection{Certified Defense With Additive Gaussian Noise.} Li  \textit{et al.} \cite{li2019certified} analyze the connection between the robustness of models against additive Gaussian noise and adversarial perturbations. They derive the certified bounds on the norm bounded adversarial perturbation, and they propose a new training strategy to improve the certified robustness. Similarly, randomized smoothing \cite{cohen2019certified} creates a smoothed classifier that correctly classifies when Gaussian noise is added to the classifier's input. Cohen  \textit{et al.} \cite{cohen2019certified} prove that this smoothed classifier can have $\ell_2$ certified robustness for an input. Both SND and the above certified defenses add Gaussian noise to the input. However, the purpose of the addition of noise in the certified defenses is to induce the classifier to gain certified robustness. Whereas SND adds noise to disturb an accurate measurement of the output to defend against query-based black-box attacks at the inference. In addition, the certified defenses use a much larger $\sigma$ ($\geq0.25$) than SND ($0.01$).

\section{Conclusion}
In this paper, we highlight that even small Gaussian input noise can effectively neutralize query-based black-box attacks and name this approach Small Noise Defense (SND). Our work suggests that query-based black-box attacks should consider the randomness of the target network as well. We demonstrate its effectiveness against eight query-based attacks with CIFAR-10 and ImageNet datasets.  Interestingly, SND is very simple and easy for defenders but difficult for attackers to bypass. SND is readily applicable to pre-trained models by adding only one code line. Due to its simplicity and effectiveness, we hope that SND will be used as a baseline of defense against query-based black-box attacks in the future. 
{\small
\bibliographystyle{ieee_fullname}
\bibliography{main}
}
\end{document}

%%%%%%%%% TITLE

%%%%%%%%% TITLE
\title{Supplementary Material: On the Effectiveness of Small Input Noise for Defending Against Query-based Black-Box Attacks}

\maketitle

\ifwacvfinal
\thispagestyle{empty}
\fi
\begin{figure*}[t]
     \centering
     \begin{subfigure}[b]{\textwidth}
         \centering
         \includegraphics[width=\textwidth]{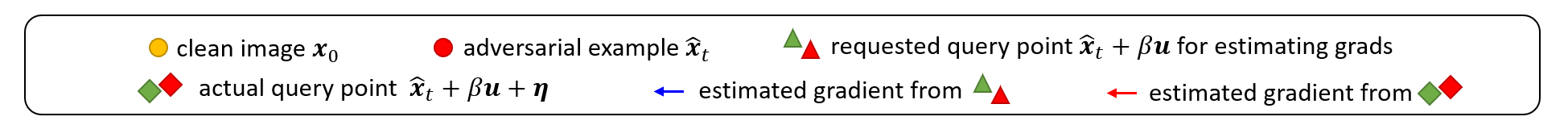}
         \label{fig:fig1_}
     \end{subfigure}
     \begin{subfigure}[b]{0.25\textwidth}
         \centering
         \includegraphics[width=\textwidth,trim={4cm 3.5cm 1.5cm 0.6cm},clip]{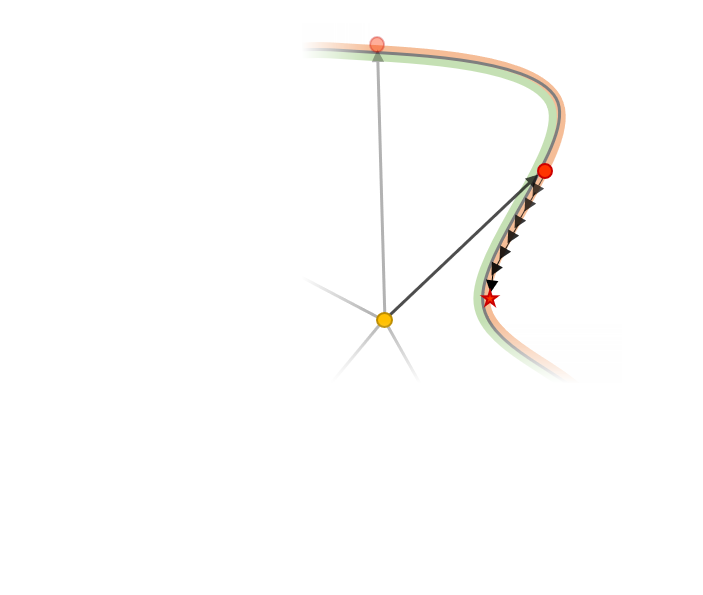}
     \end{subfigure}
          \begin{subfigure}[b]{0.25\textwidth}
         \centering
         \includegraphics[width=\textwidth,trim={4cm 3.5cm 1.5cm 0.6cm},clip]{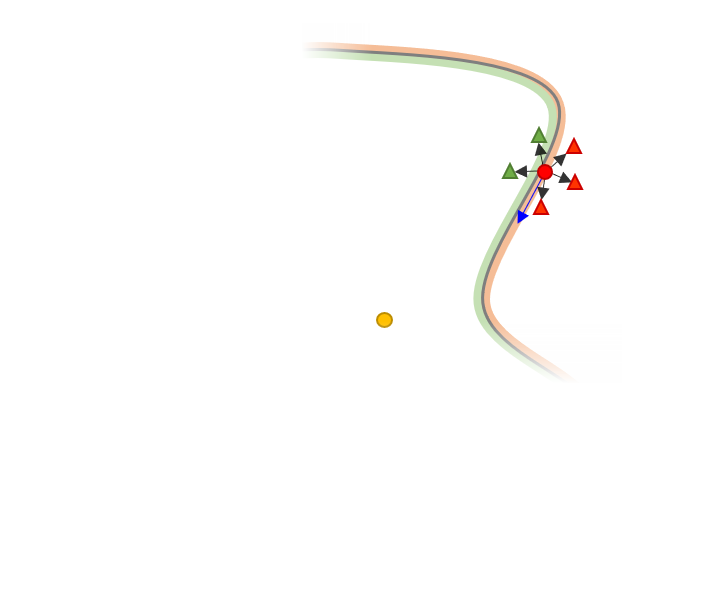}
     \end{subfigure}
      \begin{subfigure}[b]{0.25\textwidth}
         \centering
         \includegraphics[width=\textwidth,trim={4cm 3.5cm 1.5cm 0.6cm},clip]{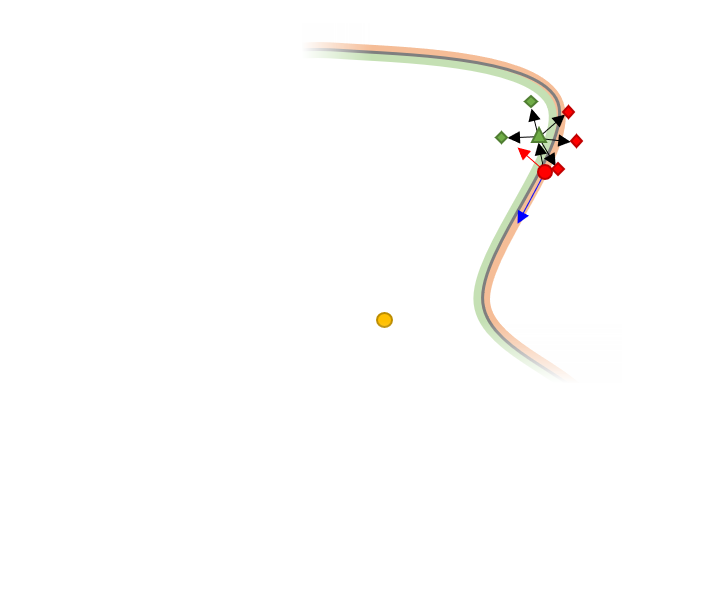}
     \end{subfigure}
        \caption{Illustrations of how small noise can defend against decision-based attacks (left) An adversary wants to reach the optimal adversarial example from an initial misclassified image. (middle) To find the next update's direction, it evaluates $\hat{\vx}_t+\beta\vu$ (right) Small noise can disturb this gradient estimation. The illustration shows that the prediction for each query image can have different values because of the small input noise.}
        \label{fig:fig2}
\end{figure*}
%%%%%%%%% BODY TEXT

\section{Illustration of Defense Against Decision-based Attack with Small Input Noise}
We illustrate the working principle of SND against decision-based attacks in Fig. \ref{fig:fig2}.

% \section{Illustrations of Defense Against Decision-based Attacks with Small Input Noise} \label{appendix:A}

\section{Detailed Experimental Settings} \label{appendix:B}

\subsection{Settings of Attack Methods}
\textbf{BA:}
We adopt BA provided by Adversarial Robustness Toolbox (ART) library \cite{art2018} with default parameters. 

\textbf{Sign-OPT:}
We take the code\footnote{https://github.com/cmhcbb/attackbox} provided by the authors without changing the special parameters of the attack.

\textbf{HSJA:}
We adopt HSJA provided by the ART library with default parameters except for increasing the maximum number of iterations to 64 to follow the authors' code\footnote{https://github.com/Jianbo-Lab/HSJA}.

\textbf{GeoDA:} We take the code\footnote{https://github.com/thisisalirah/GeoDA} provided by the authors without changing the special parameters of the attack.

\textbf{SimBA \& SimBA-DCT:}
We take the code\footnote{https://github.com/cg563/simple-blackbox-attack} provided by the authors. Following the authors, we use \verb|freq_dims|=28, \verb|order|=strided, and \verb|stride|=7 for SimBA-DCT.

\textbf{Bandit-TD:}
We take the code\footnote{https://github.com/MadryLab/blackbox-bandits} provided by the authors with default parameters except for \verb|batch_size|=1 and \verb|epsilon|=4.9.

\textbf{Subspace Attack:}
We take the code\footnote{https://github.com/ZiangYan/subspace-attack.pytorch} provided by the authors with the original settings for the $\ell_\infty$ norm untargeted attacks for the ImageNet. 
We use the pre-trained ResNet-18 and ResNet-34 trained on the \verb|imagenetv2-val| dataset as reference models that the authors provide.

\subsection{Settings of Defense Methods} 
\textbf{Baseline:} We trained a ResNet-20 model on the CIFAR-10 dataset for 200 epochs and used this model for our experiments. For the ImageNet dataset, we used the pre-trained ResNet-50 model provided by TorchVision library\footnote{https://github.com/pytorch/vision}.

\textbf{PNI:} We used the pre-trained ResNet-20 model trained on the CIFAR-10 dataset with PNI-W (channel-wise) provided by the authors.

\textbf{PGD-AT:} We used the adversarilly trained ResNet-50 model for the $\ell_2$ norm with $\epsilon_{train}=3$ provided from \textit{robustness} library \cite{robustness_lib} with PGD on the ImageNet dataset for comparisons.

\textbf{RSE:} We trained a RSE-based ResNet-20 with $\sigma_\textrm{init}=0.2$ and $\sigma_\textrm{inner}=0.1$. Considering computational efficiency, we used 5 ensembles for each prediction of RSE.

\textbf{R\&P:} 
R\&P applies random resizing and random padding to its input sequentially. It first rescales an input image of size $W\times H\times3$ with a scale factor $s$ which is sampled from $[s_{min}, s_{max}]$, and places it in a random position within an empty image of size $s_{max}W \times s_{max}Y \times 3$. Following the authors, we set $s_{min}$ and $s_{max}$ as $\frac{310}{299}$ and $\frac{331}{299}$ respectively.

\section{Evaluation of Adaptive Attacks Against SND} 
\begin{table}[t]
    \centering

        \begin{tabular}{lcccc}
        \specialrule{.1em}{.05em}{.05em} 
        Dataset &  \multicolumn{4}{c}{CIFAR-10}\\\hline
        \# of queries & 2K$\times T$ & 5K$\times T$ & 10K$\times T$ & $P_{mis}$\\\hline
        BA   ($T{=}10$) & 20.3\% & 28.5\% & 32.9\% & 0.413 \\
        SO   ($T{=}10$) & 20.5\% & 21.1\% & 21.9\% & 0.354  \\
        HSJA   ($T{=}10$) & 21.0\% & 27.3\% & 34.9\% & 0.376 \\
        GeoDA   ($T{=}10$) & 12.1\% & 12.2\% & 12.2\% & 0.413 \\
        \end{tabular}
                \begin{tabular}{lccc}
        \specialrule{.1em}{.05em}{.05em} 
        Dataset &  \multicolumn{3}{c}{ImageNet}\\\hline
        \# of queries & 5K$\times T$ & 10K$\times T$ & 20K$\times T$ \\\hline
        SimBA-DCT ($T{=}10$) & 8.8\% & 9.6\% & 11.2\%  \\
        Bandit-TD  ($T{=}10$) & 8.8\% & 8.8\% & 9.2\%   \\
        \specialrule{.1em}{.05em}{.05em} 
        \end{tabular}
                \caption{Attack success rates of the adaptive attacks against SND ($\sigma=0.01$) with $T{=}10$.}
        \label{table:adaptive}
\end{table}
We apply the expectation-based adaptive attack to BA, SO, and GeoDA on CIFAR-10 and two score-based attacks (SimBA-DCT and Bandit-TD) on ImageNet for comprehensive comparisons. The experimental results are shown in  Table \ref{table:adaptive}.

\begin{table*}[t]
\centering

\begin{tabular}{lcccccccc}
\specialrule{.1em}{.05em}{.05em} 
 & Clean Acc. (\%) & $\E||\bm{\eta}||_2$ & \multicolumn{3}{c}{Sign-OPT} & \multicolumn{3}{c}{HSJA} \\
\multicolumn{1}{c}{\# of queries} & \multicolumn{1}{c}{-} & \multicolumn{1}{c}{-} & \multicolumn{1}{c}{2K} & \multicolumn{1}{c}{5K} & \multicolumn{1}{c}{10K} & \multicolumn{1}{c}{2K} & \multicolumn{1}{c}{5K} & \multicolumn{1}{c}{10K} \\ \hline
$\sigma{=}0.001$ & 91.33 $\pm$ 0.02 & 0.055 & 20.6\% & 22.2\% & 23.4\% & 48.1\% & 67.6\% & 81.9\% \\
 $\sigma{=}0.01$ & 90.57 $\pm$ 0.09 & 0.550 & 21.7\% & 22.3\% & 22.8\% & 16.5\% & 19.9\% & 22.7\% \\
 $\sigma{=}0.01, \alpha{=}\beta{=}1$& 91.04 $\pm$ 0.12 & 0.276 & 21.6\% & 22.4\% & 22.9\% & 18.1\% & 23.9\% & 30.0\% \\
 $\sigma{=}0.01, \alpha{=}\beta{=}2$ & 91.15 $\pm$ 0.04 & 0.275 & 20.3\% & 20.8\% & 21.7\% & 19.7\% & 23.5\% & 28.5\% \\
 $\sigma{=}0.01, \alpha{=}\beta{=}0.5$ & 91.06 $\pm$ 0.06 & 0.275 & 20.9\% & 21.4\% & 22.4\% & 19.8\% & 26.3\% & 32.6\% \\
 $\sigma{=}0.02$ & 87.56 $\pm$ 0.18 & 1.098 & 26.4\%    & 26.5\% & 26.6\%  & 24.2\%   & 26.7\% & 30.1\%  \\
 $\sigma{=}0.02, \alpha{=}\beta{=}1$ & 90.17 $\pm$ 0.09 & 0.552 & 22.0\% & 22.0\% & 22.1\% & 19.6\% & 23.2\% & 25.4\% \\
 $\sigma{=}0.02, \alpha{=}\beta{=}2$ & 90.44 $\pm$ 0.08 & 0.550 & 21.6\% & 21.8\% & 22.1\% & 18.0\% & 22.1\% & 25.0\% \\
$\sigma{=}0.02, \alpha{=}\beta{=}0.5$ & 89.99 $\pm$ 0.24 & 0.549 & 22.5\% & 22.5\% & 23.3\% & 20.3\% & 24.4\% & 27.5\% \\ 
\specialrule{.1em}{.05em}{.05em} 
\end{tabular}
\caption{Experimental results of varying $\sigma$ with the CIFAR-10 dataset. We evaluate the mean and standard deviation of clean accuracy in 5 repetitive experiments on the original test dataset.}
\label{table:varying_sigma}
\end{table*}

\section{Evaluation of Varying $\sigma$ for Each Inference} 
Detailed experimental results are shown in Table \ref{table:varying_sigma}.

\begin{table*}[t]

\centering
\begin{tabular}{cccc}
 \specialrule{.1em}{.05em}{.05em} 
Attack method & \multicolumn{3}{c}{Subspace Attack} \\
\# of queries & 5K & 10K & 20K \\ \hline
Baseline & 99.6\% (99.6\%) & 100.0\% (100.0\%) & 100.0\% (100.0\%) \\
SND ($\sigma=0.01$) & 61.2\% (59.6\%) & 61.2\% (59.6\%) & 61.2\% (59.6\%) \\
SND ($\sigma=0.001$) & 64.4\% (64.4\%) & 66.4\% (66.4\%) & 68.4\% (68.4\%) \\
PGD-AT & 71.6\% (45.6\%) & 78.4\% (52.4\%) & 81.6\% (55.6\%) \\
PGD-AT + SND ($\sigma =0.01$) & 40.8\% (14.8\%) & 42.0\% (16.0\%) & 42.4\% (16.4\%) \\
PGD-AT + SND ($\sigma=0.001$) & 62.0\% (36.0\%) & 62.4\% (36.4\%) & 63.2\% (37.2\%) \\
R\&P & 73.2\% (68.4\%) & 74.0\% (69.2\%) & 74.0\% (69.2\%) \\  \specialrule{.1em}{.05em}{.05em} 
\end{tabular}
\caption{Evaluation of attack success rates of Subspace Attack against defenses on the ImageNet dataset. We also calculate the attack success rate without initially misclassified images and denote it in the parenthesis.}
\label{table:subspace_attack}
\end{table*}
\section{Evaluation of Attack Success Rates of Subspace Attack} 
Detailed experimental results are shown in Table \ref{table:subspace_attack}.

{\small
\bibliographystyle{ieee_fullname}
\bibliography{supp}
}